# Towards democratic data agency: Attitudes and concerns about online data practices

Niels Jørgen Gommesen

The Faculty of Business and Social Science, Digital Democracy Centre, Southern University of Denmark, nielsgom@sam.sdu.dk

Thursday 6th March, 2025

**Abstract**
Recent studies reveal widespread concern and increasing lack of understanding about how personal data is collected, shared, and used online without consent. This issue is compounded by limited options available for digital citizens to understand, control and manage their data flows across platforms, underscoring the need to explore how this lack of trust and transparency affects citizens' data practices including their capacities to act in a modern knowledge society. Despite the promising research within this field, important demographics are often overlooked, particularly people from marginalized social groups such as elderly, social and economically challenged communities, and younger participants. This paper addresses this gap by specifically focusing on these underrepresented groups, emphasizing the need for exploring their understandings and percepts of online data practices. Drawing on three semi-structured focus group interviews, the paper asks*: to what extent can public attitudes and concerns about data sharing on the internet inform the potential strategies and frameworks necessary to enhance digital trust and democratic data agency particularly among marginalized groups in Denmark?* The study explores the types of information, levels of transparency, and agency people desire in their daily online data practices. Additionally, it explores how these insights can potentially inform the future development of fair data strategies and technological approaches to enhance digital trust and democratic data agency. Key findings point out the need for transparent, accessible privacy policies and data management tools, emphasizing that transparency alone is insufficient without enhancing democratic agency to address trust issues and foster a more inclusive digital environment.

## 1. Introduction

In today's networked digital landscape, personal data serves as a gateway to accessing extensive details about digital citizens' private lives and daily activities. This widespread availability of personal information exposes people's most confidential details to the risk of exploitation by tech companies and third-party actors. Ruschmeier (2023) argues that the non-transparent business model of data brokers poses a significant threat to privacy, data protection and democratic values due to the lack of transparency and consent in data trading. She demonstrates that data brokers often trade personal data in a manner that conflicts with GDPR, due to issues with informed consent in digital environments (Ruschmeier, 2023, p. 28). In a recent incident concerning the misuse and sharing of personal data via Google Workspace, the Danish Data protection Agency (Datatilsynet) uncovered that public schools were sharing personal data with Google, without any legal basis (Datatilsynet, 2024). Consequently, 53 municipalities have now been instructed to ensure their data sharing practices comply with the law.

These examples highlight the complex and opaque nature of data extraction, processing and sharing, alongside a general lack of public awareness and understanding regarding these issues. While digital citizens often recognize that various platforms, connected apps and online services collect their personal information, there is a notable gap in understanding who can access their personal details, and how these data can be accessed and used by others without their knowledge or consent (Lupton, 2020). Although we may recognize that we share personal information with different services and platforms, a majority of data sharing is executed without our knowledge (Lupton, 2018) to create profiles and metadata for further distribution (Ruschmeier, 2023). This automatic exchange of information points out a critical aspect of digital interactions, where personal data seamlessly flow beyond our immediate awareness.

Participating in today's digital world involves numerous daily interactions through mobile phones, wearable technologies and mobiles apps that all generate vast amounts of behavioral data. From posting personal information on social media to using Google Maps or simply wearing our smartwatches from school and work, as digital citizens we leave data crumbs in unprecedented ways. These personal data records flow into an opaque network of corporate and state databases, where general online users have little to no control over how their data is interpreted (Cheney-Lippold, 2017). In fact, most people remain unaware of what data is collected about them and how it is used (Analyse & Tal, 2023), or how social life itself is transformed into a resource for capital (Couldry and Mejias, 2019). Instead, companies like Google and Meta leverage algorithms and user data to create a dynamic world of knowledge that contributes to shaping our lives.

In *We are Data* by Cheney-Lippold (2017), the author explores how data and algorithms increasingly shape our everyday lives. He demonstrates how these technologies shape our self-perception, digital identities, and futures, by categorizing us into different groups such as "genders", "races", "classes", and "terrorists". As a result, who we are online and what we are is tethered together and increasingly decided for us by data brokers, marketers, advertisers and tech companies. Cheney-Lippold argues that those who control data hold the power to reframe and modulate how we understand ourselves and our place in the world, "algorithmic agents make us and make the knowledges that compose us, but they do so on their own terms" (Cheney-Lippold, 2017, p. 12). The consequences of this are that we are not only losing control over our lives but also how we define it.

Kate Crawford (2021) argues that the way data is collected, understood, labelled and classified, and used to train AI systems is not merely a technical process but a social and political

intervention – a form of politics, that determines how the world is seen and evaluated (Crawford, 2021, p. 221). She highlights how the extraction of personal information raises significant concerns about privacy and reveals broader issues, where a small number of data companies wield immense power by extracting information and profits from publicly available data (see also, Zuboff, 2019). Crawford contends that AI systems are expressions of power rooted in broader economic and political processes, developed to increase profit, and empower those who use them. This underscores the profound impact these companies have on the landscape of information and economic gain, for the way artificial intelligence works in the world and which communities that are most affected.

In a digital world where personal data operates as a form of capital, the need for more transparent, actionable, and user-friendly data practices and data governance practices is more critical than ever, if we are to avoid that underserved communities are to be targets of harmful data surveillance and data extraction practices.

This paper draws on data collected from three focus group interviews with people from marginalized social groups including the elderly, underserved communities and teenagers aged 15-17. The study, explores the public's understanding of, attitudes towards, and concern about how personal data is shared and used by online platforms and services without their consent, raising the following question:

***To what extent can public attitudes and concerns about data sharing inform the potential strategies and frameworks necessary to enhance digital trust and democratic data agency particularly among marginalized groups in Denmark?***

In the following we seek to explore the following:

1: How do different social groups, particularly marginalized communities, perceive and respond to online data collection practices, and what strategies can enhance their capacity to assert data rights and agency in digital environments?
2: To what extent are transparency and accessibility in data policies sufficient, and why is fostering democratic data agency crucial for enabling meaningful control over personal data?
3: How can digital citizens be empowered to move beyond passive data acceptance and actively engage in managing and contesting their own data practices?

**2. Public understanding and perceptions of data practices**
Recent critical data studies emphasize a growing need to address the complexities and challenges associated with large scale personal data mining practices across platforms and online services, to create deeper insights into the public's understanding, engagements and concern regarding online data practices (Lupton and Michael, 2017; Kennedy *et al.*, 2020).

Key findings demonstrate that while most people are aware of the extensive collection of their personal data, they often lack sufficient knowledge to control and understand how their personal data are being shared and used by different platforms and third parties, and who have accesses to them (Lupton and Michael, 2017; Kennedy *et al.*, 2020; Analyse & Tal, 2023). A recent Danish study reveals that 82% of the data stored about us consist of unintentional or unnoticed digital traces, data that can reveal sensitive information such as health, sexual orientation, and major life changes (Analyse & Tal, 2023). A full 85% respond that they are using apps and websites because they do not believe they can realistically avoid the data collection about their person (Analyse & Tal, 2023, p. 18). 50% of the respondents answered that data collected about their person is abstract and intangible. 42% answered that they don't know about their rights to get their data deleted.

While people tend to acknowledge the value of their data for the benefits of personalized content (Lupton and Michael, 2017), a growing body of evidence highlights increasing concerns about the practices of online data collection (Kennedy *et al.*, 2021). These concerns are largely due to a pervasive lack of trust in the way personal data is accessed, used and distributed by online platforms and data services (e.g. Michael and Lupton, 2016; Kennedy *et al.*, 2020). In the Danish study by Tal and Analyse (2023), 71% of respondents expressed their greatest concern about the risk of their data being stolen or accidentally shared. Additionally, 65% of the respondents indicated significant worry about the power of big tech companies arising from the collection and sale of personal data (Analyse & Tal, 2023, p. 68). Moreover, people tended to be more concerned about data collection from private companies compared to public authorities (Analyse & Tal, 2023, p. 69).

In a separate study on public perceptions of data practices, Kennedy et al. (2020) argue that not all data practices are perceived in the same way. People experience data practies from different social positions, with social inequality playing a crucial role in shaping people's experiences of these practices (Kennedy *et al.*, 2020). Kennedy et al. (2022) argue that while we all live with our data, our experiences are shaped by our identities, who we are, and the broader social and political context. Their research found that disadvantaged or minority groups are particularly concerned about data use, often expressing worry about its impact on other marginalized groups. Furthermore, they found that concern about data usage depends on context, inequalities, and who is involved, and demographic differences play a role in the degree of concern that exists about specific data uses. Hence, they conclude that we need to be specific and precise when we talk about public concern about data uses.

The study of data practices cannot be reduced to the study of technical operations and actions, because data practices are not neutral representations of reality, but hold agentic capacities that takes part in shaping human actions (Lupton, 2018). Data practices are performative processes that carries and enacts institutional agendas, political and economic interests, cultural norms, hence to study them we also have to consider the discourses, knowledge regimes, norms and materialities that shape and inform these practices (Ruppert *et al.*, 2021, p. 34).

*From public understanding of data practices to empowerment of diverse data publics*
In the era of surveillance capitalism personal data has become a commodity at the center of the digital economy. Its infrastructure operates by extracting personal data, often without explicit consent, to shape and predict behavior for profit. Zuboff (2019) emphasizes that the right to privacy and control over our personal information and experiences are under siege, as corporations deploy extractive and predictive technologies that operate beyond our perception and control, monitoring, analysing and conditioning our behavior. These pervasive surveillance technologies influence and extend beyond individual privacy concerns, touching on broader issues of freedom and democracy. As demonstrated by Draper and Torow (2019), surveillance technologies have the power to covertly cultivate behavior on a large scale, undermining the agency of individuals and their possibilities for collective action, which are foundational to democratic governance. Zuboff (2019) states, that our inability to access and control the knowledge extracted from our experiences, has led to a shift of knowledge, power and authority towards surveillance capitalism, and left us as "exiles from our own behavior" (Zuboff, 2019, p. 100).

Kennedy and Moss (2015) argue that to enhance public agency in the digital age, we need to reimagine data mining practices in ways that can empower people to be more informed, reflective and engaged with their personal data practices. They emphasize that these changes require transparency, accountability, and regulation, including public accessibility to data and data mining tools. Kennedy and Moss suggest that data mining could be used as the means for the public to increase their knowledge about each other, reflect on matters of shared concern and improve decision-making in ways that strengthen collective actions and enable the formation of a more reflexive and active publics (Kennedy and Moss, 2015, p. 2). They stress that data mining and analytics need to be democratized in three ways. Firstly, data mining practices should be subject to greater public supervision and regulation, to address concerns about the potential negative effects of data mining of the public. However, transparency is insufficient, what is needed is accountability. This requires that data mining companies demonstrate to the public what they are doing, why they are doing it, and the effects of their actions. Secondly, data mining should be accessible and available to the public and not merely major corporations and governments. Thirdly, data mining practices should be used in ways that enable members of the public to understand each other, reflect on matters of shared concern, including how to act collectively as publics (2015, p.8). These conditions would enable data publics to exercise agency over their data and practices.

Kennedy, Poell and van Dijck (2015) emphasize the dual nature of data both as a tool for surveillance by powerful entities and a means for empowerment and resistance by the public and alternative actors. They underline the importance of thinking about agency as central for the distribution of data power, and that we understand citizen agency in relation to data structures. Moreover, they underscore the necessity of critically engaging with data and its societal implications and investigate how data and datafication affect society and potentially sustain, undermine and transform vital public values (Kennedy, Poell and van Dijck, 2015, p. 2). They argue that understanding and leveraging alternative data practices can empower individuals and communities in ways that challenge dominant power structures. Kennedy et al. write (2015), "to participate in datafied social, political, cultural and civic life, ordinary people need to understand what happens to their data, the consequences of data analysis, and the ways in which data-driven operations affect us all" (2015, p. 6).

In line with this Ruppert et al. 2021 argue that data citizens requires opportunities for making their rights claim and the possibility to contest and intervene in their own subjectivation and take part in how data is made about them and the communities of which they are a part (Ruppert *et al.*, 2021, p. 291). Ruppert (2019) proposes that one way forward for democratizing personal data could be to engage data citizens in the co-production of the data about themselves. This includes their digital actions, interactions, how their experiences are systematically categorized, included or excluded (Ruppert, 2019, p. 636). These examples emphasize the authors' call for a more nuanced and democratic approach to datafication, emphasizing the potential for data to serve public interests and enhance their data agencies.

In the following, we use the concept of *democratic data agency* to refer to practices that enable a diverse public to actively engage in the datafied dimensions of social, political, and civic life. Agency is a central quality of the concept that suggests new possible new ways for citizens to claim their rights and to contest and intervene in their own data subjectivation. This includes new possibilities for contesting and regulating who accesses our data, increased transparency and accountability in how data is collected and used, and public accessibility that enables collective discussion and critique of matters of public concern. Hence, data becomes a tool for empowering diverse publics, leveraging alternative data practices that empowers diverse individuals and communities.

This study's emphasis on democratic data agency is especially crucial in light of ongoing challenges related to data exploitation, unclear data practices, and limited public awareness. Non-transparent data practices threaten privacy and democratic values, particularly for marginalized communities, making the need for transparent, user-friendly data governance practices more urgent than ever. Its novelty lies in its inclusion of diverse participant groups—elderly participants, underserved communities, and teenagers—offering unique insights into the varying levels of understanding, concerns, and vulnerabilities across social groups. By researching these varied experiences, the study emphasizes the need for more inclusive, transparent, and accessible data governance strategies. This research highlights that transparency alone is insufficient; empowering digital citizens requires providing them with the tools, knowledge, and agency to manage their data proactively. By advocating for inclusive, user-centered data governance frameworks, this study contributes significantly to the broader discourse on data rights, emphasizing the importance of enabling all digital citizens to contest, shape, and navigate the complex digital landscape actively.

## 3. Methodology

In this study, we draw on the empirical material from three focus group interviews (n=24), with 8 participants per group. The focus group interviews made it possible to collectively explore and discuss the different subjects' experiences, knowledge, and concerns about online data practices and privacy measures. The focus groups were conducted in person between November 2023 and February 2024 and lasted 90 minutes on average. To ensure confidentiality, the participant names were anonymized and assigned identifiers (participants 1-24).

| Identifiers | Age | Gender | Demographic category | Concerned about private life |
|---|---|---|---|---|
| P1 | 61 | F | Underserved participant | Concerned |
| P2 | 60 | M | Underserved participant | Predominantly concerned |
| P3 | 58 | F | Underserved participant | Concerned |
| P4 | 54 | M | Underserved participant | Predominantly concerned |
| P5 | 41 | M | Underserved participant | Predominantly concerned |
| P6 | 31 | F | Underserved participant | Concerned |
| P7 | 56 | M | Underserved participant | Concerned |
| P8 | 37 | F | Underserved participant | Predominantly concerned |
| P9 | 71 | F | Elder participant | Concerned |
| P10 | 69 | F | Elder participant | Dont know |
| P11 | 80 | M | Elder participant | Predominantly concerned |
| P12 | 72 | M | Elder participant | Concerned |
| P13 | 71 | M | Elder participant | Concerned |
| P14 | 69 | M | Elder participant | Predominantly concerned |
| P15 | 74 | F | Elder participant | Predominantly concerned |
| P16 | 67 | F | Elder participant | Predominantly concerned |
| P17 | 15 | F | Young participant | Unconcerned |
| P18 | 15 | F | Young participant | Unconcerned |
| P19 | 15 | M | Young participant | Unconcerned |

| P20 | 16 | M | Young participant | Unconcerned |
| P21 | 16 | M | Young participant | Unconcerned |
| P22 | 15 | M | Young participant | Unconcerned |
| P23 | 15 | M | Young participant | Unconcerned |
| P24 | 16 | M | Young participant | Unconcerned |

*table 1: social demographic factors of the participants*

The interview guide was organized into four themes. The first explored citizens' data practices and online behaviours. The second examined their knowledge and use of privacy settings. The third addressed trust and transparency regarding the collection and sharing of personal information. Finally, the fourth theme addressed citizens' concerns and attitudes towards online data collection. The interview concluded with follow-up questions about the participants' interests and willingness to learn more about online data privacy and data sharing/extraction.

For the recruitment of the focus group participants, we used the recruitment company Kanter to ensure a screening process that made it possible for us to connect with more diverse social groups such as elders 65+ years of age, social and economically underserved groups, including teenagers 15-17 years of age, with an equal selection of men and women for each group. In terms of the general requirements for the three groups, the participants should be general users of the Internet and online services, in a way that makes it possible for them to engage in a discussion about the everyday online data practices and attitudes and concerns towards online data extraction. In the screening process managed by Kanter, people were asked about their gender, age, background (e.g. financial support, daily allowance, or early retirement), internet use, expertise, and concern about data extraction and their understanding and attitudes towards data privacy.

The elderly group included people 65+ years of age, who represent a segment of the population that is believed to have limited exposure to and understanding of digital technologies and data practices compared to younger generations. The second focus group included people from socially and economically underserved communities who are believed to face unique challenges when it comes to data privacy, and who may be more vulnerable to data exploitation, discrimination, and exclusion. With this group, we expected to reveal insights on how socio-economic status may significantly impact people's data practices, data privacy concerns and their understanding of data privacy. The third group involved teenagers 15-17 years of age, who are believed to have a strong online presence. With this group, we expect to get a deeper understanding of their daily data practices experiences and concerns regarding data extraction and data privacy. By exploring the unique perspectives, attitudes, and concerns towards data practices across the three demographic groups, we aimed to shed light on how data practices impact marginalized communities and compare these to the general Danish population, to potentially reveal inequalities in data privacy awareness and possible needs for creating more inclusive, accessible, transparent online data practices that people can engage with and evaluate.

**4. Data analysis**

In our study, our focus groups were conducted in the professional and controlled environment provided by Kantar, which encouraged open and candid dialogue among the participants. To ensure a comprehensive documentation, the sessions were recorded using both video and audio, which allowed us to capture verbal responses but also non-verbal expressions and body language.

This approach enriched our dataset, by providing deeper contextual insights for our analysis.

Following the data collection, we initiated a qualitative thematic analysis (Braun and Clarke, 2006, 2019). Initially, the focus group recordings were transcribed verbatim, transforming the rich, dynamic discussions into text for detailed explorations. This transcription formed the foundation for our emergent data analysis approach, allowing patterns and themes to surface naturally rather than being constrained by pre-existing hypotheses.

We commenced with an open coding process, systematically organizing the data into manageable segments. Each segment related to identified patterns or themes emerging from our initial readings of the transcribed text. This iterative process involved deep immersion into the data, with the researcher carefully documenting initial codes that captured key points or concepts discussed by the participants. These codes were then aggregated into potential themes that encapsulated the essence of our research questions. At this stage, redundant codes were discarded, and related codes were refined and combined into coherent themes, enduring a clear and meaningful representation of the data.

Hereafter, we reviewed the emerging themes to verify their alignments with the collected data. This process sometimes led to the consolidation of similar themes or the division of broader themes into more nuanced sub-themes. These refinements were crucial in ensuring that the final themes accurately reflected the data and contributed meaningfully to our research objectives.

The final stage in our analysis involved defining and naming the themes. We conducted a detailed analysis of each theme, supporting our interpretations with direct quotes from the focus group transcripts. This approach grounded our thematic analysis in the participants' actual discussions, vividly illustrating how the themes emerged from their perspectives and experiences. By gathering rich data through focus groups and systematically identifying and analyzing key themes, we gained valuable insights into citizens' understanding of and attitudes toward data practices.

## 5. Findings

In the following, we analyse people's perspectives and understandings of online data practices, including their knowledge, attitudes, concerns, and trust across three social groups. We address potential challenges arising from these insights and propose key findings that suggest new approaches for *democratic data agency*.

- Theme 1: Digital citizens' knowledge and understanding of online data practices
- Theme 2: Attitudes towards and concern about online data collection
- Theme 3: Trust and transparency
- Theme 4: Conditioned data subjects and Digital resignation

**Theme 1: Digital citizens' knowledge and understanding of online data practices**
Our interviews revealed that social groups perceive online data sharing differently. While older generations tend to be more attentive to and concerned about sharing their personal information and data with different online actors, younger participants tend to pay less attention to their daily data practices.

P17 (15 years, F) tells us: "about my attitude towards data sharing, it´s something I think about when I am online, and 'maybe I should not share that', however I don't deeply care about it either. I give it a brief thought and then decide whether it really matters". P18 (15 years, F) aligns with this:

"So, it's not something I think too much about. It's something I think about sometimes, and especially if it's a dangerous website". P19 (15 years, M): "I do not really think about data sharing", and P20 (16 years, M): "I usually just skip it if there are cookies and stuff. I don't think too much about it". And finally, P21 (16 years, M): "Accepting cookies has become so common that I now accept all cookies without much thought".

When asked if people read privacy policies, P17 (15 years, F) states that she often does not read them: "I would probably say that if it's a website I trust, I'm probably more likely to skim through it. But if it's a sketchy website, I'm just considering whether I should say no thanks. I don't read it through." P20 (15 years, M) explains that the information he receives is often overwhelming: "Yes. Most of the time, they bombard you with a lot of information all at once, which you don't want to scroll through. Even if you do scroll through it, you often don't manage to read it. It doesn't feel manageable to sit down and read everything when you want information quickly".

Our findings demonstrate that school students often ignore data-sharing policies and cookie settings, and do not think too much about where their data ends up. At the same time, older generations are more attuned to but also more concerned about their sharing of personal information. P9 (71 years, F) tells us: "I am extremely uneasy about the traces I leave behind when I search the web. And I think it is completely unclear where they end up. You can see that if you have searched for one or the other, all sorts of ads suddenly pop up". P9 (71 years, F) is concerned about the data traces she leaves behind, she believes that it is unclear what happens to her data and that it is too demanding to read the several pages of policies to get to the content. Another participant P16 (67 years, F), also finds data sharing hard to comprehend:

> I honestly find it a little hard to relate to, somehow. Because I believe that some of the things you collect can be beneficial. For example, research about diseases etc. and at the same time, it's like, what are they going to use it for [personal information]? So I think it's a bit difficult to take a stand on. I have an easy-going attitude, but there's a sense of Big Brother lurking.

The examples demonstrate that digital citizens often find it challenging to understand online data collection. On one hand, data collection can be useful and benefit society, but on the other people do not know what happens to their data.

Another participant, P12 (72 years, M), who has worked in IT for many years, tells us: "I have started to be concerned". While referring to a news article he read about Internet users being monitored far more extensively than previously known. He tells us: "and then one becomes even more worried". P13 (71 years, M, former database administrator), is in line with this and adds: " In my opinion, it is alarming if you are not careful. When making inquiries of various kinds, one should always figure out how to clean up after oneself when finished". P13 consistently prioritizes data privacy and is cautious about sharing personal information and as he tells us, always considers "to remove traces, cookies and things that you know will be collected". Also, P9 (71 years, F) is deleting her digital footprints and information that could be used to track her online activities:

> I understand that one might feel a bit more secure knowing that they can remove their search history and cookies from Google. I regularly remove my search history and cookies, but I still see things that I never searched for or wanted to see. But that is just how it is, it seems like a self-imposed security measure.

Karen protects her online identity by removing cookies and online search history. However, she is skeptical about the effectiveness of her efforts and has come to terms with the fact that data sharing is an inevitable aspect of using online services, there is little she can do to prevent it.

Our key findings demonstrate that not just younger generations, but also older ones tend to skip reading details such as cookie policies on websites. As P15 (74 years, F) tells us: "I am happy when a website only asks for necessary cookies. Then I accept it. Otherwise, if I have to go into settings and such, I cannot be bothered. Then I click 'allow all' so I can proceed." P15 does not read the cookie policies but argues that all websites should only have the 'necessary cookies option' available. Also, P12 (72 years, M) is not reading the policies carefully, "I skim them, but I do not read them thoroughly".

Participants from the underserved group have varied perceptions of data extraction. Joachim expresses resignation when it comes to preventing sharing his personal information: "I am probably more resigned. Just saying, well that is how it is! I have accepted that things are the way they are. Because…Otherwise, how the hell can you control all of it? There is not one central place where you can easily access and see it". P6 (31 years, F, underserved participant) experiences something similar, however, she is also more concerned about it: "Like P7 (56 years, underserved participant), I feel a bit the same, that is the way things are, But of course, once in a while. I am a little worried about it. Among other things, with Facebook and Instagram right now, they are probably selling things [personal data]". P6 has come to terms with the inevitability of online data extraction, feeling there is little she can do to prevent it. At times, she worries about the possibility of her personal information being sold. Another participant, P4 (54 years, M), states that: "[data collection] is a necessary evil. You get what you get", suggesting that he has accepted that data extraction is just a natural part of using the Internet. P8 (37 years, F, underserved participant), tells us that she attempts to read privacy policies but often the text is too complicated to understand: "I've tried to read it. But I don't think I understand it. No. It is not that kind of everyday reading". Hence, the lack of transparency in personal data sharing is not the only issue at stake. The policies are also too complex and can be overwhelming to read for the underserved group.

**Theme 2: Attitudes towards and concern about online data collection**

Our findings show that the elder participant group possesses the highest level of knowledge about online data practices, but they are also the ones who are most critical towards it. P9, for instance, asserts that online data extraction by private companies is fundamentally undemocratic:

> *I think it's deeply undemocratic that as a citizen and consumer, you do not know where your data ends up*. In an ideal world, there could be a place where you could say, *"Yes, you may use it for anything," "No, you may only use it for this," or "You may not use it at all."* But *I don't believe that's realistic*. Because the companies behind it all are so powerful. And in my opinion, they are *beyond ordinary democratic control*.

This example highlights peoples' concerns about the lack of transparency and control over personal data flows. P9 expresses her frustration with the current situation, emphasizing the need for a different system that provides users with more agency and control regarding the collection and usage of their data. However, they doubt the feasibility of such an alternative due to the power held by the data companies, suggesting that these entities operate beyond the reach of traditional democratic oversight.

However, when it comes to sharing personal information with public institutions people are more willing to share their data, as P16 tells us:

> I must confess that when it comes to my health, especially after undergoing some hospital procedures and whatnot, and regarding my medical records, I trust that my data is being taken care of. I will have to, otherwise it would be impossible to keep track of it all.

This example demonstrates that P16 is more willing to share personal data with public institutions. She emphasizes the importance of trusting the system's ability to manage her health data, as navigating the healthcare system without this trust would be difficult. Hence, the example reflects a broader concern regarding data privacy and the necessity for people to have confidence in public institutions' handling of their sensitive information. Also P15 (74 years, F) tells us that she is confident sharing her data with public institutions: "Do I trust that they uphold that policy, or do I not trust it? I do. And to a greater extent in public institutions, etc., which have a conscious democratic oversight, unlike private companies." P15 argues that she is confident in sharing her private information with public institutions because there are no commercial interests involved. This resonates with Karens opinion:

> When it comes to private companies, they are naturally driven by the need to generate profit, which makes me wary of how they might use my personal information. Although I do not fully trust the public sector, I believe they are less likely to misuse my data for commercial gain.

This example demonstrates that participants are more careful with and more concerned about sharing their personal information with private companies who potentially sell their information to third parties, compared to democratic institutions that are not motivated by commercial interests. Moreover, when it comes to sharing personal information, the context seems to matter to the individual users, as P8 tells us:

> I probably don't think too much about it. If a product or service is located in Denmark compared to abroad, I am more inclined to be careful. If it's in Denmark, then I don't worry as much because I believe they have better control over it.

Hence, as the example tells us people are less concerned about sharing their private data with Danish companies compared to services located outside Denmark. However, it matters what kind of information people are sharing, if the information is sensitive such as Cpr. Numbers, people are more careful and will only sharing with trusted companies and institutions such as Apoteket.dk and when making appointments with their doctor.

**Theme 3: Trust and transparency**

In our interviews with the participants across the three social groups, we often hear that people are somewhat concerned about the extensive online data extraction because often it is not transparent what happens to their data and how it is used by others. As P9 (71 years, F) tells us:

> I find it concerning and undemocratic that I don't know where the things I search for, whether in my official capacity or privately, end up. It's undemocratic that I lack control over where the information I search for, whether related to purchasing or opinions and politics, comes from or ends up. I would like to have control over that myself.

In this example, P9 finds it concerning and undemocratic that she does not have more control over what happens to her data flows. For her, it is crucial to have the capability to make her claims right and be able to decide if others can use her data. P9 states:

> So, if I want to express myself about some current political issue, I can do it in various ways. I can write a post. I can attend relevant meetings and so on. Then *I've made the decision myself that here I want my opinions to be known to others. Or what I want to know more about to be known to others.* I don't have that when I search the internet. I have no idea where they end up. I have no idea what they're being used for.

People want more agency in terms of what happens to their online data flows, what personal information is shared and who can use them. They want more control over their personal information, opinions, and the decision-making behind the online data flows. In other words, they want the ability as digital citizens to make their rights claim and decide what is being shared with others.

People tell us that it is often difficult to comprehend not only privacy settings and cookie settings but also what content they are looking at, and if the website is a safe place to share their information. P8 (37 years, F, underserved participant), tells us about an episode where she got scammed online via email, from a fake website:

> I mean, you hardly even know what you're looking at. Because I recently got scammed. Because I booked a hotel stay on Booking.com. And then I receive an email with a link, and confirmation codes and everything. I completely felt for it and it turned out to be fraud.

The example illustrates that it can be difficult for people to not only understand online policies and privacy settings but also to determine if a website is a secure place to share their private information. This challenge arises from a lack of transparency.

Another participant, P6 (31 years, F, underserved participant), suggests that she does not know to whom she is giving her data:

> Well, I don't know who I am giving my data to. It is that thing again with cookies. You just click "yes" because you rarely have the time and energy to read it all. So, I am not sure who I am giving my data to.

The example suggests that people often give their information away and hope for the best because either there is no transparency on the sites they are using, or they do not have the energy and time to read the extended privacy policies (add younger generations' perspectives).

In our discussion with the school students, we often hear that people lack sufficient knowledge about what happens to their data and how it is used and shared e.g. by third parties. As the two students P24 and P18 tell us, P24: "No, I didn't even know it was happening, so this is new information" and P18: "It's not something you are told. I just can't see where I would get that information from". The example highlights that school students often lack sufficient knowledge about what happens to their online data and how personal information is shared with third parties. Moreover, and particularly for the younger generations, there seems to be a tendency of decreased curiosity and concern regarding their online data sharing.

**Theme 4: Conditioned data subjects and Digital resignation**
In our discussions with the participants across the three social groups we heard that people often do not know what happens to their data. In our conversations with the elders, we frequently hear that people cannot foresee the consequences of their data-sharing practices, how their data is used including how to manage their privacy settings. As P15 (74 years, F) tells us:

> I want to configure my browser and settings to decrease ads, but I often struggle to grasp the consequences of my choices when prompted to accept various options. Once, I tried to be very restrictive and found out that I could not use some of the functions I enjoy. It made it difficult for me to get a clear overview. I need to focus on the overall picture before diving into the details.

In this example, we hear that even though people want to protect their privacy and limit their data sharing, they often struggle to understand the implications of their choices, e.g. when configuring browser settings to reduce advertising. The implications of this are that they miss the overall perspective, which makes it even more difficult to navigate privacy settings and manage specific details. Others like P1 (61, years, underserved), however, argue that they do not think much about privacy settings in their everyday life: "No, actually, I do not think much about it in daily life. I am also one of those who do not look through all 10 pages". Similarly, younger generations argue that they also do not think much about privacy settings. P17 (15 years, F), tells us that privacy settings are not something she is conscious about when visiting a website:

> Well, it's not something I think you see a lot about; I mean, it comes up sometimes when you visit a website, but not something you think much about. No. It's just something you expect to be appropriate in a way.

In general younger generations are inclined to be less concerned about sharing their data, they do not read privacy policies, and in general, have more trust in the online services they use. As P18 (15 years, F) and P17 (15 years, F) tell us:

> P18: "There are many apps and websites that you just trust and think, 'Yeah, that's fine.'"

> P17: "And it's also a bit like "everyone uses it". Like, for example, Instagram. Everyone has Instagram, after all. So you're probably more inclined to trust them."

These findings suggest that especially the younger generations are less inclined to manage their privacy and cookies settings, and in general are less concerned about what happens to their data, because often they do not understand the implications of their online data practices and how to limit data extraction. As P17 tells us:

> I know too little to bother doing anything about it. I do not really know what happens when I click 'accept' for cookies, or things like that. I do not really know where it could lead, so I do not bother thinking about it.

Our findings demonstrate that younger generations are generally less interested in managing privacy and cookie settings. This is often because they do not think about it, and also because they may not fully understand the implications of their online choices, due to a lack of transparency regarding their data.

As P17 suggests: "I think maybe it's like they don't really show you what's happening, or tell you what you're saying yes to. You sort of associate it with something they're trying to hide away". We heard the same from the elders about a lack of transparency and comprehension of data sharing and privacy policies. P6 (31 years, F, underserved) shared that she does not know who she is sharing her data with. She often accepts the cookies because she does not have the energy to read through everything and simply wants to move on. Many people do not know who receives their data because often it is too complicated to track their own data sharing and understand its implications. As a result, some tend to accept that data sharing is just a natural part of using the internet, beyond their control and agency. Joachim compares using online services with services in a physical shop:

> No, but it's like this: "Welcome to the store. If you don't want to talk to me, I can't help you." So you have to choose to share information to a degree that you also get proper service. And that's fair enough. If you think you can go into a physical store with a sheet over your head and remain unknown, well, then they can't help you or meet your needs. And it's the same online. I actually think it's okay. It's not a concern. It's just a premise.

In this example we also hear that some people have accepted that personal data sharing is necessary if you want to use an online service, even though they know the premise of sharing their information, hence they are not concerned about sharing their private information because they have accepted its terms of usage. For the younger generations, we hear something similar, however, in this situation, people have less knowledge about what happens to their data, and what they can do to understand and change the circumstances. As P19 (15 years, M) and P20 (16 years, M) tell us:

> P19: I feel like it's something you can't really do anything about, or maybe you can, but I don't really know what you can do, so I just don't think about it, I guess.

> P20: And it's not like I'm a big celebrity or anything, so I don't think about it at all. It's just me, it's not like I'm someone they can get a lot of info from and use for much.

These examples demonstrate that people perceive data sharing from very different perspectives and have varied attitudes towards what happens to their data. Moreover, they lack the capacity to control what happens to their personal information. People react very differently to data sharing. Some people are deeply concerned about companies collecting their data and trying to protect themselves, but they often lack the knowledge to do so effectively. Others see data sharing as a necessary part of navigating the Internet, as they want functions and services tailored to their interests. However, across the three social groups, there is a general sense of confusion and helplessness about what happens to their data and how they can effectively manage and navigate the data privacy landscape.

The lack of transparency in online services' data policies, combined with complicated documentation and vague data-sharing practices, leaves ordinary users feeling like they have lost control of their data. They struggle to understand and manage how their data is used and end up feeling powerless to change the situation. As a result, they often merely accept that their data is being collected and used because they do not know how to prevent it or how their data is used.

In summary, the absence of data agencies affects online users' actions, thoughts, decisions, and opinions about their data, placing it out of their control. This leads to the ordinary data citizens being limited in their abilities to assert their rights and contest the status quo. As a result, people passively accept their roles as data subjects, being exploited for the agendas of various companies.

**6. Discussion: Towards democratic data agencies**
To participate in a datafied social, political, cultural and civic life, regular folks need to understand what happens to their data, the consequences of data extraction and analysis and how they affect us all (Cheney-Lippold, 2017). However, as we have demonstrated above few people have the knowledge, skills and energy to understand and act on their online data practices. Moreover, people do not believe that they are in control of their personal data, rather they have resigned from acting on and securing their sharing of personal information. These results suggest that ordinary data citizens are in urgent need of new possibilities for making their rights claim and the possibility to contextualize and intervene in their own subjectivation.

Key findings demonstrate that different social groups not only act differently online but also exhibit varying levels of understanding and attitudes towards online data practices. While older participants tend to be more attentive and concerned about the consequences of data sharing, younger participants are less so, often accepting cookie policies without considering future implications. Younger participants also rarely read privacy policies, finding them overwhelming and difficult to process. This automatic exchange of information points out critical aspects of our digital interactions, where personal data flows beyond our immediate awareness (Lupton, 2020). The greater concern among the older participants shows that they are more likely to take protective measures, such as managing their data-sharing habits and cleaning their online activities to maintain privacy. Despite generational differences, there are some commonalities as both the younger and older participants often skip through the overly complicated and long data sharing policies and cookie settings. This points to a broader issue of accessibility and clarity in privacy policies across age groups. Participants from underserved groups also highlighted the complexity and opacity of data-sharing practices, often expressing resignation about current data extraction practices. They feel there is little they can do to prevent it, a condition referred to as digital resignation by Draper and Turow (2019). This resignation cultivates confusion and indecision, leaving individuals feeling powerless to control their online information despite the desire to do so. These findings underscore the diverse attitudes and levels of understanding regarding online data practices across different age groups. Younger participants' casual approach may make them more vulnerable to privacy risks, while older participants' concerns highlight the need for clearer and more manageable data policies. The broader societal issue of inaccessible and overly complex privacy policies affects not only older participants, who may struggle with digital literacy, but also younger and underserved participants who lack the tools or knowledge to engage with these policies meaningfully.

Our findings reveal a nuanced landscape of attitudes towards online data extraction, with significant variation across participant groups. The attitudes and concerns center on trust, transparency and perceived control over personal data. Older participants demonstrated the highest levels of knowledge and capabilities about online data practices, coupled with the most critical attitudes. Their call for a system where users can control their data usage highlights a desire for greater agency and transparency over personal data. However, participants also expressed skepticism about the feasibility of such a system due to the significant power wielded by data companies, which they perceive as operating beyond democratic oversight. This skepticism and concern were particularly pronounced among older participants, who voiced concerns about the opaque nature of data flows and the potential misuse of personal information by private companies. These concerns stem from a fundamental mistrust in the motivations of these companies, seen as prioritizing profit over public interest. Across all participant groups, there was a general willingness to share personal data with public institutions.

People's trust in public institutions is rooted in necessity and the belief that these entities manage data more responsibly, operating under democratic oversight and being less likely to misuse personal information for commercial gain. Furthermore, the type of data being shared also influences people's levels of concern. Sensitive information such as personal identification numbers elicits greater caution, with participants preferring to share such data only with trusted entities like healthcare providers and pharmacies. This reflects a more nuanced understanding of risk, based on both the sensitivity of the data and the trustworthiness of the receiving party. These findings underscore the complexity of attitudes towards online data collection. The high level of concern among the older participants, particularly towards private companies, highlight a need for greater transparency and control mechanisms. Their trust in public institutions suggests that policies emphasizing public oversight and accountability could enhance trust and willingness to share data. The geographical and contextual elements of trust identified among the underserved participants suggest that localized regulatory frameworks and clearer guidelines on data practices could mitigate concerns about data sharing. Ensuring that companies provide easily understandable privacy policies and clear information about data usage could address the accessibility issues highlighted by participants.

Kennedy and Moss (2015) suggest that enhancing public data agency, data mining practices should be subject to greater public supervision and regulation, to address potential concerns about the negative effects of data mining on the public. In line with this Ruppert et al. (2021) argue that data citizens require opportunities for making their rights claim and the possibility to contest and intervene in their own subjectivation and take part in how data is made about them and the communities of which they are a part (Ruppert et al., 2021, p. 291).

Our interviews with the participants groups reveal a pervasive concern about the lack of transparency in online data practices, significantly impacting trust in digital services and the entities collecting personal data. Participants consistently expressed unease regarding how their data is used. Older participants, in particular, emphasized the need for greater agency in managing their personal data, underscoring the importance of deciding how their information, opinions and political expressions are shared and used. For many the lack of transparency in online data flows was seen as a significant barrier to exercising such control and found undemocratic by some participants.

The demand for greater control over personal data is a recurring theme in our findings. Participants expressed a desire to manage how their data is used and accessed. Without this control, people tend to feel disempowered and distrustful within digital environments. Moreover, participants reported difficulties understanding privacy settings and assessing the security of online services. The complexity and opacity of these settings contribute to this challenge, undermining trust in online transactions and platforms. Among the underserved participants, it was noted that they frequently consent to data sharing without fully understanding the implications, due to the cumbersome nature of the policies. This lack of transparency and clarity further erodes trust, leaving users resigned to inevitable data extraction. Additionally, discussions with school students revealed a significant gap in their understanding of online data collection. Many students lack adequate knowledge about how their personal data is shared with third parties, showing diminished curiosity and concern about data sharing, which may increase their vulnerability to privacy issues. These findings underscore the critical need for improved transparency in data practices to build trust among digital citizens and emphasize the demand for mechanisms that allow active data management. Simplifying privacy policies and enhancing education around data practices, particularly for younger users, could help mitigate the identified trust issues.

Participants across the groups consistently report difficulties in understanding the *consequences* of their personal data-sharing practices, including how to *manage* their privacy settings. These experiences underscore a broader issue that while users want to protect their privacy, they often lack the knowledge or tools to manage their settings effectively which potentially can lead to a sense of frustration and resignation towards changing their privacy settings. Particularly among underserved groups, and school students, there is a noted disinterest in what happens to their data, likely stemming from a general lack of understanding about online data practices. Many students are unaware of the consequences of sharing their data and accepting cookies, leading to passive acceptance of these practices as a natural part of using online services.

Participants, particularly from underserved groups and younger generations, indicate that privacy settings are not a major concern in their daily lives. This suggests a general lack of engagement with privacy settings among younger users, who tend to trust the platforms they use and are thus less inclined to question online data practices. School students tend to have higher levels of trust in online services and are less skeptical towards the apps and services they use, such as Instagram, especially if these are used among their peers.

This indicates social conditioning, where the ubiquity and popularity of certain platforms foster a sense of inherent trust that potentially reduces concerns about data privacy. Although participants have different attitudes toward data sharing, there is a shared concern about companies collecting their data, with some participants actively taking steps to protect their personal information. However, often people lack the knowledge to do so effectively. Others view data sharing as a natural part of using the Internet because they want functions and services tailored to their interests. Across the three social groups, there is a general sense of confusion and helplessness about what happens to their data and how they can effectively manage and navigate the data privacy landscape.

Key findings demonstrate that the lack of transparency in online services' data policies, overly complicated documentation, and abstract data sharing practices make users feel like they are losing control of their data. Not only do they struggle to understand and manage their data, but they also end up feeling they lack the agency to change the current situation. The absence of data agencies affects online users' actions, thoughts, decisions, and opinions about their data, placing it out of their control. As a result, ordinary digital citizens merely accept that their data is being collected and used because they do not know how to prevent it or how their data is used, leading people to passively accept their roles as data subjects, exploited for the agendas of various companies.

In addressing the significant challenges identified in the digital landscape, such as the pervasive lack of transparency and complexity in data practices that overwhelm users, it becomes evident that enhancing transparency alone is not sufficient. To truly empower users, especially diverse and underserved communities, we must promote democratic data agency: a concept that refers to practices that enable a diverse public to actively engage in the datafied dimensions of social, political, and civic life, enhancing their ability to claim rights, context, and intervene in their own data subjectivation (see Ruppert *et al.*, 2021, p. 221 for more on subjectivation).

Democratic data agency emphasizes the necessity for increased transparency and accountability in how data is collected and used but also advocates for public accessibility that allows for collective discussion and critique of matters of public concern. By leveraging alternative data practices, this approach seeks to empower diverse individuals and communities, enabling them to use data as a tool for empowerment. Thus, while user-friendly privacy management tools are crucial, they must be part of a broader strategy that includes fostering capabilities and knowledge among digital citizens. This strategy should aim to transform data into a tool that supports active and informed participation in managing personal data and contesting data practices. Such a holistic approach is critical for asserting rights effectively and reshaping the data landscape to support democratic engagement and agency among all users.